\documentclass[pra,twocolumn,showpacs,letterpaper,showpacs,superscriptaddress]{revtex4}
\usepackage{amsfonts}
\usepackage{amsmath}
\usepackage{amssymb}
\usepackage{graphicx}
\newcommand{\be}{\begin{equation}}
\newcommand{\ee}{\end{equation}}
\newcommand{\bea}{\begin{eqnarray}}
\newcommand{\eea}{\end{eqnarray}}

\preprint{ }

\begin{document}

\title{Casimir energy between media-separated cylinders: the scalar case}

\author{F. C. Lombardo}
\affiliation{Departamento de F\'\i sica {\it Juan Jos\'e
Giambiagi}, FCEyN UBA, Facultad de Ciencias Exactas y Naturales,
Ciudad Universitaria, Pabell\' on I, 1428 Buenos Aires, Argentina}

\author{F. D. Mazzitelli}
\affiliation{Departamento de F\'\i sica {\it Juan Jos\'e
Giambiagi}, FCEyN UBA, Facultad de Ciencias Exactas y Naturales,
Ciudad Universitaria, Pabell\' on I, 1428 Buenos Aires, Argentina}

\author{P. I. Villar}
\affiliation{Computer Applications on Science and Engineering Department,
Barcelona Supercomputing Center (BSC),
29, Jordi Girona 08034 Barcelona, Spain}

\author{D. A. R. Dalvit}
\affiliation{Theoretical Division, MS B213, Los Alamos National Laboratory, Los Alamos, NM 87545, USA}

\date{today}

\begin{abstract}
We derive exact expressions for the Casimir scalar interaction energy between media-separated eccentric
dielectric cylinders and for the media-separated cylinder-plane geometry using
a mode-summation approach. Similarly to the electromagnetic Casimir-Lifshitz interaction energy between
fluid-separated planar plates, the force between cylinders is attractive or repulsive depending on the
relative values of the permittivities of the three intervening media.
\end{abstract}


\maketitle


\section{Introduction}

The sign of van der Waals and Casimir \cite{Casimir} forces between
media-separated plates can be tailored by carefully choosing the
frequency-dependent permittivities of the plates and the intervening
media. For the simplest geometry of two planar slabs separated by a
fluid, the Lifshitz theory \cite{Lifshitz} predicts a repulsive
force when the permittivity of the fluid is intermediate between
those of the two slabs for a large range of frequency; otherwise,
the force is attractive. Such repulsive fluctuating forces in fluids
have been measured recently \cite{repulsion-fluids}. In order to go
beyond the simple parallel plate geometry several approximation
methods have been employed, including the proximity force
approximation (PFA) \cite{Derjaguin}, the pairwise summation
approach (PWS) \cite{Parsegian}, and dilute-limit expansions
\cite{Milton,Golestanian}. Exact approaches have been developed to
treat complex geometries, including semi-analytical approaches 
based on scattering theory
\cite{Emig,Lambrecht,Klich,Messina}, and on the mode-summation technique
combined with the Cauchy theorem \cite{ouroldwork, numerical}.
There are also fully
numerical methods based on Green functions \cite{Johnson}, 
worldline approaches \cite{Gies}, and the combination 
of  boundary methods
traditionally used to compute eigenvalues of the Helmholtz equation
with the Cauchy  theorem in order to perform
the sum over modes \cite{pmm}.
Most of these computations have
been performed for vacuum-separated dielectric or metallic plates.
Fluid-separated complex geometries have been the subject of recent
works, including a numerical study of Casimir repulsive forces and
torques between fluid-separated eccentric cylinders
\cite{RodriguezPRL} (see also \cite{jamal} 
for the analysis of the Casimir force in a configuration consisting of an 
object contained inside a spheroidal cavity filled with a dielectric medium).

Here we derive the exact analytical expression for the Casimir
interaction in media-separated cylindrical configurations, including
eccentric cylinders and the cylinder-plane geometry. For simplicity,
we consider in this work the case of a quantum real scalar field
satisfying usual boundary conditions on the interfaces. Our approach is a 
generalization
of our previous work on the electromagnetic Casimir interaction in
cylindrical geometries with perfect reflectors \cite{ouroldwork} to
the case of dielectric media, and is based on the computation of the
Casimir energy as a sum of the zero-point eigenfrequencies of the
three-media geometry. We should note that this approach requires the
notion of {\sl real} zero-point energies, and therefore the three
media should have negligible absorption in the whole range of
frequencies relevant for the Casimir interaction. The case of
absorbing media can be considered with alternative techniques
\cite{paperwithMilonni,Francescopaper}.


\section{Mode-summation approach: Eccentric cylinders}

We consider a massless scalar field $\phi$ in the presence of two parallel, eccentric 
dielectric cylinders of radii $a$ and $b$ ($a<b$) and length $L \gg a, b$ separated by a 
fluid. The eccentricity (i.e. the distance
between the centers of the cylinders) will be denoted by $a \delta$, with
$\delta$ a dimensionless number. The inner and outer cylinders have permittivity
$\epsilon_1(\omega)$ and $\epsilon_3(\omega)$ respectively, and $\epsilon_2(\omega)$ is
the permittivity of the fluid.  In order to enclose the system in a finite volume, 
we will include a very large cylinder of radius $R\gg a,b$, concentric with the outer cylinder. 
We will assume that the scalar field satisfies Dirichlet boundary conditions on the larger cylinder.
The limit $R\to\infty$ will be taken at the end of the calculation.

Using the translational symmetry along the $z$ direction, the solutions
of the corresponding Klein-Gordon equation can be written as
$\phi({\bf r}, z, t) = \varphi({\bf r}) e^{- i \omega t + i k_z z}$,
where ${\bf r}=x {\hat{\bf x}} + y {\hat{\bf y}}$.
For the region inside the inner cylinder ($0<r<a$), it is 
convenient to use a polar coordinate system
${\bf r} = (r, \theta)$ centered on the inner cylinder. The 
most general solution
for $\varphi({\bf r})$ in this region is
\begin{equation}
\varphi(r, \theta) = \sum_n C_n J_n(\lambda_1 r) e^{i n \theta},
\end{equation}
where $C_n$ are constants to be determined. For the region outside
the cylinder of radius $b$ it is convenient to use a coordinate system ${\bf
r} = (\rho, \phi)$ centered on it. The solution  in
this region ($b < \rho < R$) has the form
\begin{equation}
\varphi(\rho, \phi) = \sum_n \left[\bar{C}_n J_n(\lambda_3 \rho) +
\bar{D}_n H^{(1)}(\lambda_3 \rho)\right] e^{i n \phi} ,
\label{zone3}
\end{equation}
where $\bar{C}_n, \bar{D}_n$ are constants to be determined. Finally, in the region
between the cylinders where the fluid is located, the solution is
\begin{equation}
\varphi(r, \theta) = \sum_n \left[A_n J_n(\lambda_2 r) + B_n
H^{(1)}_n(\lambda_2 r)\right] e^{i n \theta},
\end{equation}
written in the $(r, \theta)$ coordinate system, or
\begin{equation}
 \varphi(\rho, \phi) = \sum_n \left[\bar{A}_n J_n(\lambda_2 \rho) +
\bar{B}_n H^{(1)}(\lambda_2 \rho)\right]e^{i n \phi},
\end{equation}
written in the $(\rho, \phi)$  coordinate system. In these equations
we have defined $\lambda^2_i = \epsilon_i(\omega) \omega^2/c^2 - k_z^2$ ($i=1,2,3$).

Imposing the continuity of the field and its derivative at the
interface between the inner cylinder and the fluid ($r=a$) one
obtains
\begin{eqnarray}
C_n J_n(\lambda_1 a) &=& A_n J_n(\lambda_2 a) + B_n
H_n^{(1)}(\lambda_2 a) , \nonumber  \\
\lambda_1 C_n J_n'(\lambda_1 a) &=&\lambda_2 \left[ A_n
J_n'(\lambda_2 a) + B_n H_n'^{(1)}(\lambda_2 a)\right].
\label{bca}
\end{eqnarray}
Therefore $B_n = -  A_n [ J_n(\lambda_2 a) / H_n^{(1)}
(\lambda_2 a)] R^{(1-2)}_n$,
where
\begin{equation}
R^{(1-2)}_n= R^{(1-2)}_n(k_z, \omega) =  \frac{1- \frac{\lambda_2}
{\lambda_1}\frac{J_n(\lambda_1 a)
J_n'(\lambda_2 a)}{J_n(\lambda_2 a)J_n'(\lambda_1 a)}}{1-
\frac{\lambda_2}{\lambda_1}\frac{J_n(\lambda_1 a)
H_n'^{(1)}(\lambda_2 a)}{J_n'(\lambda_1 a)H_n^{(1)}(\lambda_2 a)}}.
\end{equation}

Before imposing the boundary conditions at $\rho = b$, we consider
that at $\rho= R \gg a, b$ the scalar field satisfies Dirichlet 
boundary conditions
\begin{equation}
\bar{C}_n J_n(\lambda_3 R) + \bar{D}_n H_n^{(1)}(\lambda_3 R) = 0.
\end{equation}
The final results do not depend on whether we impose
Dirichlet or Neumann boundary conditions at $\rho=R$. As we know
from previous calculations \cite{ouroldwork}, in order to evaluate
the Casimir energy it is convenient  to rotate to imaginary
frequencies $\omega\to i \xi$, which implies that, in the above
equation,  $J_n(\lambda_3 R) \to I_n(\tilde\lambda_3 R)$ and
$H_n^{(1)}(\lambda_3 R) \to K_n(\tilde\lambda_3 R)$, where
$\tilde\lambda_i=\sqrt{\epsilon_i(i \xi) \xi^2/c^2+k_z^2}$. Taking
into account the behavior of the modified Bessel functions $I_n$ and
$K_n$ for large arguments, it follows that the coefficients
$\bar{C}_n$ must vanish when $R\to\infty$. An equivalent procedure
would be, without enclosing the system in a finite volume, to
consider only outgoing waves in Eq.(\ref{zone3}), which amounts to
imposing $\bar{C}_n=0$. Either way,  the boundary conditions at
$\rho = b$ have the form
\begin{eqnarray}
\bar{D}_n H_n^{(1)}(\lambda_3 b) &=& \bar{A}_n J_n(\lambda_2 b)
 + \bar{B}_n H^{(1)}_n(\lambda_2 b) , \nonumber \\
\frac{\lambda_3}{\lambda_2} \bar{D}_n H_n'^{(1)}(\lambda_3 b) 
&=& \bar{A}_n J'_n(\lambda_2 b) +
\bar{B}_n H_n'^{(1)}(\lambda_2 b).\label{bcb}
\end{eqnarray}
It is then possible to find a relation between $\bar{B}_n$ and 
$\bar{A}_n$ as
$\bar{B}_n = - \bar{A}_n [J_n(\lambda_2 b) / H_n^{(1)}(\lambda_2 b) ]  
R^{(2-3)}_n$,  where
\begin{equation}
R^{(2-3)}_n= R^{(2-3)}_n(k_z, \omega) =
\frac{1- \frac{\lambda_2}{\lambda_3}\frac{J_n'(\lambda_2 b)
H_n^{(1)}(\lambda_3 b)}{J_n(\lambda_2 b)H_n'^{(1)}(\lambda_3 b)}}{1-
\frac{\lambda_2}{\lambda_3}\frac{H_n^{(1)}(\lambda_3 b)
H_n'^{(1)}(\lambda_2 b)}{H_n'^{(1)}(\lambda_3 b)H_n^{(1)}(\lambda_2 b)}}.
\end{equation}

The coefficients associated with the solution written with coordinates
centered at the inner cylinder can be related to those centered at the 
outer cylinder
by the use of the addition theorem for Bessel functions
\begin{eqnarray}
A_n &=& \sum_m \bar{A}_m J_{n-m}(\lambda_2 a \delta) ,\nonumber \\
B_n &=& \sum_m \bar{B}_m J_{n-m}(\lambda_2 a \delta ).
\label{addition}
\end{eqnarray}
Combining equations (\ref{bca}), (\ref{bcb}),  and (\ref{addition}) 
one obtains a
linear, homogeneous system of equations. The solution of this system
is non-trivial only if det$[{\bf M}]= 0$, where
\begin{equation}
{\bf M}_{nm} = \left[
1 - \frac{R^{(1-2)}_n}{R^{(2-3)}_n} ~\frac{H_m^{(1)}(\lambda_2 b) 
J_n(\lambda_2 a)}{J_m(\lambda_2 b)
H_n^{(1)}(\lambda_2 a)}
\right]
\;J_{n-m}(\lambda_2 a \delta) .
\label{matrixM}
\end{equation}

The interaction energy between the inner (media 1) and outer (media 3)  
cylinders is given 
by the sum of the zero-point eigen-energies
that are the solutions to det$[{\bf M}]= 0$. We shall assume that the 
absortion in the media is negligible and therefore the zero-point 
eigen-frequencies will be real. Moreover, as we are enclosing the system 
in a big cylinder of radius $R$, the solutions to det$[{\bf M}]= 0$ 
form a discrete set $\omega_n(k_z)$, $n = 1, 2, 3, ...$. In order to compute the Casimir interaction 
energy, we will use Cauchy's theorem which involves an analytic 
continuation of the determinant to the complex plane. As the matrix elements 
in Eq.(\ref{matrixM}) are functions of $\lambda_i= \sqrt{\epsilon_i(\omega) \omega^2/c^2 - k_z^2}$, 
there will be branch points at $\lambda_i =0$. For example, if the permittivities 
are described using the plasma model $\epsilon_i(\omega) = 1 - \omega_{\rm P_i}/\omega^2$, 
the branch points are located on the real axis at
$\omega_i = \pm  \sqrt{c^2 k_z^2 + \omega_{\rm P_i}^2}$, where 
$\omega_{\rm P_i}$ is the plasma frequency of the medium ${\rm i}$. The presence of branch points
is typical for geometries with translational invariance, even for the case of a single cylinder (see for instance
Refs. \cite{nesterenko,nesterenko2}).

The Casimir interaction energy $E_{13}$ is  

\begin{equation}
E_{13} = \frac{\hbar c L}{4\pi} \int_{-\infty}^{+\infty}dk_z \sum_n \left(\omega_n(k_z)- \omega^\infty_n(k_z)\right),
\end{equation}
where $ \omega^\infty_n(k_z)$ are the solutions of det${\bf M}_\infty= 0$ with ${\bf M}_\infty$  
given by Eq.(\ref{matrixM}) with $b$ replaced by $b_\infty \gg b$.

\begin{figure}[ht]
\centering
\includegraphics[width=5cm]{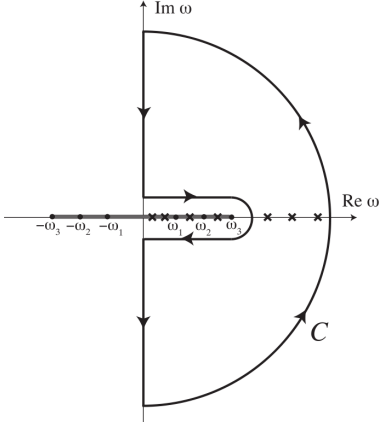}
\caption{Integration contour in the complex plane. Dots indicate the different 
branch points, where we have set (without loss of generality) $\omega_1<\omega_2<\omega_3$. 
Crosses denotes the eigen-frequencies. There is a branch cut in the interval 
$[-\omega_3,\omega_3]$ on the real axis.} \label{fig0}
\end{figure}

Let us denote as $\omega_{\rm max}$ the maximun between $\omega_1$, $\omega_2$, and $\omega_3$, the 
positions of the branch points. We split the sum over eigen-frequencies as 

\begin{equation}
 \sum_n \omega_n(k_z) = \sum_n \omega^<_n(k_z) + \sum_n \omega^>_n(k_z),
\end{equation}
where $\omega^<_n$ ($\omega^>_n$) are the eigen-frequencies smaller (bigger) than 
$\omega_{\rm max}$. The sum over $\omega^>_n$ can be written as

\begin{equation}
 \sum_n \omega^>_n(k_z) = \frac{1}{2\pi i}\oint_{C} dz\, z 
\frac{d}{dz}\ln[\det{\bf M}]\, ,
\end{equation}
where $C$ is the contour shown in Fig.\ref{fig0}. The contour $C$ starts at $i\infty$ 
goes along the positive imaginary axis to the origin, circumvents the three branch points, 
follows the negative imaginary axis up to $-i\infty$, and 
closes with a large semi-circunference.

One can show that the contribution to the integral of the semi-circunference vanishes. Moreover, 
the contribution of the segments above and below the real axis gives minus the sum over 
$\omega^<_n(k_z)$. Therefore, we end up with a representation of the interaction energy as 
an integral on the imaginary axis

\begin{equation}
E_{13}=\frac{\hbar c L}{8\pi^2 i}\int_{-\infty}^\infty dk_z\int dz\, z 
\frac{d}{dz}\ln[\det{\bf M}/\det({\bf M}_\infty)].
\end{equation}

From this point onwards, the calculation of the exact Casimir 
interaction energy between media-separated material
cylinders proceeds as in the case of vacuum-separated perfectly 
reflecting cylinders \cite{ouroldwork}. The main
difference with that calculation is the presence of $R_n^{(1-2)}$ and 
$R_n^{(2-3)}$, both of which are equal to unity in the
perfectly reflecting case. Following the same procedure as in 
\cite{ouroldwork},
the interaction energy can be written as

\begin{equation}
 E_{13}=\frac{\hbar c L}{4 \pi^2 }\int_{-\infty}^\infty dk_z 
\int_{0}^\infty d\xi \, \ln\det ({\bf I}- {\bf A})\,\, ,
\label{exact13}
\end{equation}
where the matrix elements of ${\bf A}$ are
\begin{eqnarray}
A_{np} &=& \tilde{R}^{(1-2)}_n \frac{I_n(\tilde \lambda_2 a)}
{K_n(\tilde\lambda_2 a)}  \label{matrixelements}\\
&\times & \sum_m \frac{K_m(\tilde\lambda_2 b)}{I_m(\tilde\lambda_2 b)} 
\frac{1}{\tilde{R}^{(2-3)}_n}
I_{n-m} (\tilde\lambda_2 a \delta) I_{p-m}(\tilde\lambda_2 a\delta).
\nonumber
\label{Anp13}
\end{eqnarray}
The functions $\tilde{R}_n$ are the analytic continuation of the functions $R_n$
to  imaginary frequencies $i\xi$:
\begin{equation}
\tilde{R}^{(1-2)}_n = R^{(1-2)}_n(k_z, i \xi) = \frac{1- \frac{\tilde\lambda_2}
{\tilde\lambda_1}\frac{I_n(\tilde\lambda_1 a)
I_n'(\tilde\lambda_2 a)}{I_n(\tilde\lambda_2 a)I_n'(\tilde\lambda_1 a)}}{1-
\frac{\tilde\lambda_2}{\tilde\lambda_1}\frac{I_n(\tilde\lambda_1 a)
K_n'(\tilde\lambda_2 a)}{I_n'(\tilde\lambda_1 a)K_n(\tilde\lambda_2 a)}} ,
\label{elR}
\end{equation}
and
\begin{equation}
\tilde{R}^{(2-3)}_n = R^{(2-3)}_n(k_z, i \xi) = \frac{1- \frac{\tilde\lambda_2}
{\tilde\lambda_3}\frac{I_n'(\tilde\lambda_2 b)
K_n(\tilde\lambda_3 b)}{I_n(\tilde\lambda_2 b)K_n'(\tilde\lambda_3 b)}}{1-
\frac{\tilde\lambda_2}{\tilde\lambda_3}\frac{K_n(\tilde\lambda_3 b)
K_n'(\tilde\lambda_2 b)}{K_n'(\tilde\lambda_3 b)K_n(\tilde\lambda_2 b)}}.
\label{elRtilde}
\end{equation}

In order to derive this result we enclosed the system into a large 
cylinder, alternatively one can deal with the unbounded configuration following 
the approach described in Ref.\cite{nesterenko}. 

\subsection{Attraction-repulsion crossover}

In this subsection we study the condition for the crossover from attractive to repulsive
interaction. For simplicity, we concentrate ourselves on the case of concentric cylinders.
In the particular case of $\delta = 0$ the matrix $A_{np}$ becomes
diagonal, and the Casimir interaction energy reduces to
\begin{eqnarray}
E_{13}&=&\frac{\hbar c L}{4\pi^2 }\sum_n \int_{-\infty}^\infty dk_z \int_{0}^\infty d\xi \, \nonumber \\
&\times & \ln\left[ 1 -
\frac{\tilde{R}^{(1-2)}_n }{\tilde{R}^{(2-3)}_n} \frac{I_n(\tilde{\lambda}_2 a)}{K_n(\tilde{\lambda}_2 a)}
\frac{K_n(\tilde{\lambda}_2 b)}{I_n(\tilde{\lambda}_2 b)}\right].
\label{concen}
\end{eqnarray}
We now analyze the signs of $\tilde{R}^{(1-2)}_n$ and 
$\tilde{R}^{(2-3)}_n$ as a function of the three
permittivities $\epsilon_i$.
For the $\tilde{R}^{(1-2)}_n$ coefficients, we rewrite Eq.(\ref{elR}) 
using the new variables $x \equiv \tilde{\lambda}_2 a$ and
$\tilde{\lambda}_{12} \equiv \tilde{\lambda}_1/\tilde{\lambda}_2$, obtaining
\begin{equation}
\tilde{R}^{(1-2)}_n = \frac{1-\frac{1}{\tilde{\lambda}_{12}}\frac{ I_n(\tilde{\lambda}_{12} x
)I'_n(x)}{I_n(x) I'_n(\tilde{\lambda}_{12} x )}}{1-\frac{1}{\tilde{\lambda}_{12}}
\frac{ I_n(\tilde{\lambda}_{12} x)K'_n(x)}{K_n(x) I'_n(\tilde{\lambda}_{12} x )}}
\label{elR2}.
\end{equation}
Since $I_n(x)$, $K_n(x)$, and $I'_n(x)$ are always positive 
for $x>0$, while $K'_n(x)$ is always negative,
the denominator in  Eq.(\ref{elR2}) is positive. Therefore, 
the sign of $\tilde{R}^{(1-2)}_n$ is
determined by the numerator $N$, that can be written as 

\begin{equation}
 N = 1 - \frac{f(x)}{f(\lambda_{12}x)},
\end{equation}
where $f(x) = x \frac{I_n'(x)}{I_n(x)}$. It is 
possible to show \cite{bessels} that $f(x)$ is an increasing 
function of $x$. Therefore $f(x) > f(\tilde{\lambda}_{12} x)$ 
for $\tilde{\lambda}_{12} < 1$ and viceversa when $\tilde{\lambda}_{12} > 1$. 
This implies that $\tilde{R}^{(1-2)}_n>0$ when 
$\epsilon_1 > \epsilon_2$ ($\tilde{\lambda}_{12} > 1$), and
$\tilde{R}^{(1-2)}_n < 0$ when
$\epsilon_1 < \epsilon_2$ ($\tilde{\lambda}_{12} < 1$).

Using a similar argument it is possible to analyze 
the sign of $\tilde{R}^{(2-3)}_n$ as given by Eq.(\ref{elRtilde}). 
The numerator is always positive, and then the
sign of $\tilde{R}^{(2-3)}_n$ is governed by the denominator $D$ which reads

\begin{equation} D = 1 - \frac{g(\tilde{\lambda}_{23}u)}{g(u)},
\end{equation}
where $u = \tilde{\lambda}_3 b$, $\tilde{\lambda}_{23} = \tilde{\lambda}_2/\tilde{\lambda}_3$, 
and $g(u) = - u \frac{K_n'(u)}{K_n(u)}$. As $g(u)$ is an increasing function, it is easy to check that
$\tilde{R}^{(2-3)}_n < 0$ for  $\epsilon_2 > \epsilon_3$, 
and $\tilde{R}^{(2-3)}_n > 0$ for $\epsilon_2 < \epsilon_3$.

The sign of the Casimir interaction energy $E_{13}$ in
Eq.(\ref{concen}) is determined by the sign of the ratio
$\tilde{R}^{(1-2)}_n/\tilde{R}^{(2-3)}_n$. When this ratio is
positive (negative) the energy is negative (positive). 
From the above considerations it follows that for $\epsilon_1 <
\epsilon_2 < \epsilon_3$ (corresponding to $\tilde{R}^{(1-2)}_n>0$
and $\tilde{R}^{(2-3)}_n<0$), or for  $\epsilon_1 >\epsilon_2 >
\epsilon_3$ (corresponding to $\tilde{R}^{(1-2)}_n<0$ and
$\tilde{R}^{(2-3)}_n>0$), the Casimir pressure is repulsive. For all
other cases the Casimir pressure is attractive. Strictly speaking, 
the sign of the energy is not enough to determine the atractive/repulsive 
character of the pressure. However, we have checked using simple 
numerical evaluations that the integrand of Eq.(\ref{concen}) is always a monotonous 
function of $b$.

As we will see
in Section IV, when the cylinders
are eccentric ($\delta \neq 0$), the same conditions imply a
repulsive or attractive force, with the concentric configuration
being an equilibrium situation. When $\epsilon_1 < \epsilon_2 <
\epsilon_3$  or when $\epsilon_1 >\epsilon_2 > \epsilon_3$ this
equilibrium is stable, and it is unstable in all other situations.
Just as in the case of the electromagnetic Casimir-Lifshitz interaction between
media-separated planar slabs, it is only necessary that the above
mentioned inequalities between the three different permittivities
should hold in the relevant range of frequencies in
Eq.(\ref{concen}). This frequency range is determined by the
geometrical parameters, particularly by the minimum distance between
the cylinders. In cases where the three permittivities satisfy the
inequalities needed for repulsion in some frequency range, but
violate them in some other frequency range, the global sign of the
force results from a competition between the different contributions
to the integrand in Eq.(\ref{concen}) (see \cite{RodriguezPRL} for
further details).


\subsection{Perfect conductivity limit}

In this subsection we  study the perfect conducting
limit of the expression for the interaction energy of 
Eq.(\ref{exact13}). For the scalar field,  ``perfect
conductivity" corresponds to Dirichlet boundary conditions on
the interfaces. These conditions can be formally achieved
for large values of the permitivities $\epsilon_1$ and 
$\epsilon_3$ of the two cylinders.

Using the asymptotic expansions for $I_n(x)$, $K_n(x)$, 
and their derivatives, it follows that
in the limit $\tilde{\lambda}_1 \rightarrow \infty$
the function $\tilde{R}^{(1-2)}_n$ takes the form
\begin{equation}
\tilde{R}^{(1-2)}_n   \approx
1 - \frac{\tilde{\lambda}_2}{\tilde{\lambda}_1} \left[ 
\frac{I'_n(\tilde{\lambda}_2 a)}{I_n(\tilde{\lambda}_2 a)}
- \frac{K'_n(\tilde{\lambda}_2 a)}{K_n(\tilde{\lambda}_2 a)} 
\right].
\label{rapprox}
\end{equation}
Similarly, in the limit $\tilde{\lambda}_3 \rightarrow 
\infty$ the function $\tilde{R}^{(2-3)}_m$ takes the form
\begin{equation}
\tilde{R}^{(2-3)}_n   \approx
1 + \frac{\tilde{\lambda}_2}{\tilde{\lambda}_3} \left[ 
\frac{I'_n(\tilde{\lambda}_2 b)}{I_n(\tilde{\lambda}_2 b)}
- \frac{K'_n(\tilde{\lambda}_2b)}{K_n(\tilde{\lambda}_2 b)} 
\right].
\end{equation}
In this perfect-conductors limit, the matrix elements of 
${\bf A}$ can be written as $A_{np} \approx A_{np}^{\rm PC}
- \Delta A_{np}$, where
$A_{np}^{\rm PC}$ are the matrix elements corresponding 
to taking the perfect conductor (PC) limit $\tilde{R}^{(1-2)}_n =
 \tilde{R}^{(2-3)}_n=1$ (see \cite{ouroldwork}), and
\begin{eqnarray}
&& \Delta A_{np} = \frac{\tilde{\lambda}_2 }{{\tilde\lambda}_1 }
\left[ \frac{I'_n(\tilde{\lambda}_2 a)}{I_n(\tilde{\lambda}_2 a)} 
- \frac{K'_n(\tilde{\lambda}_2 a)}{K_n(\tilde{\lambda}_2 a)} \right]
\nonumber \\
&& \times
\frac{I_n(\tilde \lambda_2 a)}{K_n(\tilde\lambda_2 a)}
\sum_m \frac{K_m(\tilde\lambda_2 b)}{I_m(\tilde\lambda_2 b)} 
I_{n-m} (\tilde\lambda_2 a\delta) I_{p-m}(\tilde\lambda_2 a\delta )
\nonumber \\
&& +  \frac{\tilde{\lambda}_2 }{{\tilde\lambda}_3} 
 \frac{I_n(\tilde \lambda_2 a)}{K_n(\tilde\lambda_2 a)}
\sum_m
\left[ \frac{I'_m(\tilde{\lambda}_2 b)}{I_m(\tilde{\lambda}_2 b)} - 
\frac{K'_m(\tilde{\lambda}_2 b)}{K_m(\tilde{\lambda}_2 b)}
\right] \nonumber \\
&& \times
\frac{K_m(\tilde\lambda_2 b)}{I_m(\tilde\lambda_2 b)} I_{n-m}
 (\tilde\lambda_2 a\delta) I_{p-m}(\tilde\lambda_2 a\delta) .
\end{eqnarray}
The Casimir interaction energy is therefore 
$E_{13} \approx E_{13}^{\rm PC} + \Delta E_{13}$, where
\begin{eqnarray}
\Delta E_{13} &=& - \frac{\hbar c L}{4 \pi^2} \int_{-\infty}^{\infty} 
dk_z \int_0^{\infty} d\xi
\; {\rm Tr} \; \frac{\Delta {\bf A}}{1 - {\bf A}^{\rm PC}} .
\end{eqnarray}

In this limit, and assuming that the $\epsilon_i$ are constants (i.e., no dispersion), it is possible to obtain the 
explicit dependence of the interaction energy
$\Delta E_{13}$ on the three permittivities. The key point is to note that 
\begin{eqnarray}
\Delta E_{13} &=&  \frac{\hbar c L}{4 \pi^2} \int_{-\infty}^{\infty} 
dk_z \int_0^{\infty} d\xi\left( \frac{1}{\tilde\lambda_1}f_1(\tilde\lambda_2,a,b,\delta)
\right .\nonumber\\
&+& \left. \frac{1}{\tilde\lambda_3}f_2(\tilde\lambda_2,a,b,\delta)\right)\,\, ,
\label{key}
\end{eqnarray}
for some functions $f_i$. We change variables in the integral above introducing 
polar coordinates $(\eta,\varphi)$ in the plane
$(\sqrt{\epsilon_2}\xi / c, k_z)$, so that $\tilde\lambda_2=\eta$ and 
$\tilde\lambda_i=\eta\sqrt{(\epsilon_i/\epsilon_2 - 1)\cos^2\varphi+1}$ for
$i=1,3$. The integral in $\varphi$ can be computed explicitly and gives
\begin{equation}
\int_0^{2\pi}\frac{d\varphi}{\sqrt{(\epsilon_i/\epsilon_2 - 1)\cos^2\varphi+1}}
\approx\sqrt{\frac{\epsilon_2}{\epsilon_i}}\ln\left(\frac{\epsilon_i}{\epsilon_2}\right)
\end{equation}
for $\epsilon_1, \epsilon_3 \rightarrow \infty$.
Inserting this result in Eq.(\ref{key}) we obtain
\begin{equation}
\frac {\Delta E_{13}} {L}\approx
\frac{\ln(\epsilon_1/\epsilon_2)}{\sqrt{\epsilon_1}} G_1(a,b,\delta)
+\frac{\ln(\epsilon_3/\epsilon_2)}{\sqrt{\epsilon_3}} G_2(a,b,\delta)\, ,
\label{largeeps1}
\end{equation} 
where the functions $G_i$ involve integrals in the radial coordinate $\eta$.

In previous  works \cite{Milton}, the Casimir energy was evaluated 
using dilute-limit expansions. Eq.(\ref{largeeps1}) gives the dependence
with the permittivities in the opposite limit. The result shows that the 
convergence to the case of perfect conductivity is rather slow. 


\section{Cylinder-plane geometry}

In the case of perfect conductors, we have shown in \cite{ouroldwork} that the
cylinder-plane configuration is contained as a particular case of
the exact formula for eccentric cylinders. In this Section we obtain the matrix elements for the cylinder-plane configuration from
the media-separated cylinders Eqs. (\ref{exact13}) and
(\ref{matrixelements}). As in \cite{ouroldwork}, let us
consider a cylinder of radius $a$ above an infinite plane.
The permittivities are $\epsilon_1$ inside the cylinder,  $\epsilon_2$ between the cylinder and
the plane,  and $\epsilon_3$ in the region below the plane.
Let us denote by $H$ the distance between the center of the cylinder
and the plane. The expression for  the interaction energy in the media-separated cylinder-plane geometry can be obtained from 
the eccentric cylinders formula Eq. (\ref{exact13})
taking  the limit $b/a$, $\delta\rightarrow
\infty$, keeping $H/a = b/a - \delta$ fixed. Using the asymptotic
limit of the Bessel functions, it is possible to show that the
coefficient ${\tilde R}_n^{(2-3)} \rightarrow \frac{{\tilde
\lambda}_3 + {\tilde \lambda}_2 }{{\tilde \lambda}_3 - {\tilde
\lambda}_2 }$,  and therefore it can be taken outside the sum in
Eq. (\ref{matrixelements}). After this, and making use of the uniform
expansion and the addition theorem of Bessel functions, it can be
also shown that \cite{ouroldwork}
\begin{eqnarray}
 &&\sum_m \frac{K_m(y + l)}{I_m(y + l)} I_{m - n}(y) I_{m - p}(y)\nonumber \\
& \approx & \sum_m K_m (y + l) I_{m - p - n}(y - l) =
K_{n + p}(2 l) ,
\end{eqnarray}
in the limit $y \rightarrow \infty$.
Finally,  the expression for the matrix elements in the 
cylinder-plane geometry is given by
\begin{equation}
 A_{np}^{\rm cp}= {\tilde R}_n^{(1-2)}
\frac{{\tilde \lambda}_3 - {\tilde \lambda}_2}{{\tilde \lambda}_3 
+ {\tilde \lambda}_2} \frac{I_n({\tilde \lambda}_2 a)}
{K_n({\tilde \lambda}_2 a)} K_{n+p}(2{\tilde \lambda}_2 H) .
\label{cylinder-plane} 
\end{equation} 
This generalizes the result of \cite{cil-plane} for vacuum-separated, perfectly conducting cylinder-plane to the three-media cylinder-plane geometry. 

The sign of the Casimir
interaction energy for the cylinder-plane geometry is ruled by
the signs of ${\tilde R}_n^{(1-2)}$ and $({\tilde \lambda}_3 -
{\tilde \lambda}_2)/({\tilde \lambda}_3 + {\tilde \lambda}_2)$. The
attractive-repulsive character of the force
depends on the relative values of $\epsilon_1$, $\epsilon_2$
and $\epsilon_3$, in the same way as for the eccentric cylinders.

The dependence of the Casimir energy on the permittivities in the
perfect conductivity limit can also be  derived from
Eq.(\ref{cylinder-plane}).  The procedure is similar to the one used
to obtain Eq.(\ref{largeeps1}), so we only quote the final result

\begin{equation}
\frac{\Delta E_{13}^{\rm cp}}{L}\approx 
\frac{\ln(\epsilon_1/\epsilon_2)}{\sqrt{\epsilon_1}} G^{\rm cp}_1(a,d)
+\frac{\ln(\epsilon_3/\epsilon_2)}{\sqrt{\epsilon_3}} G^{\rm cp}_2(a,d)\, ,
\label{cylinder-plane-large-eps1}
\end{equation} 
for some functions $G^{\rm cp}_i$.
In the next Section we will provide numerical evaluations
that confirm this behavior.


\section{Numerical evaluations}

In this Section, we show numerical results for the Casimir
interaction energy between eccentric media separated cylinders and a
cylinder in front of an infinite plane. For simplicity, we consider
the dispersion-less case in which $\epsilon_i$ are constants.

The value of the Casimir interaction energy is obtained by the numerical
evaluation of Eq.(\ref{exact13}), through the use of the different definitions
of the matrix elements $A_{np}$ depending upon the geometry considered
(Eqs.(\ref{Anp13}) and (\ref{cylinder-plane})). We numerically
compute the Casimir interaction energy using a FORTRAN
program which defines the matrix elements of $\mathbf{A}$, 
computes the corresponding eigenvalues. and finally
performs the frequency and wavevector integrations. The
parameters used by the program are  the dimension 
of the ${\bf M}$ matrix, the number of addends 
corresponding to each element of the ${\bf M}$ matrix, the integration
limits ${k_z}_{\rm max}$ and $\xi_{\rm max}$, 
and the desired precision. In the following,
we show the numerical results obtained.

\begin{figure}[ht]
\centering
\includegraphics[width=8.9cm]{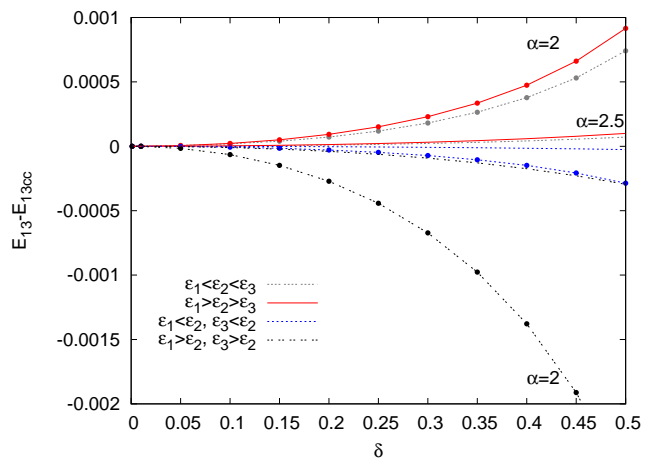}
\caption{(Color online). Exact Casimir interaction energy difference between the
eccentric and concentric configurations as a function of $\delta$
for different values of $\alpha = b/a$. 
Curves with circles correspond to $\alpha=2.0$, while those without circles correspond to $\alpha=2.5$. 
In all cases $\epsilon_1=2$, $\epsilon_2=5$ and $\epsilon_3=50$,
ordered as indicated in the legend. Energy is plotted in units of $\hbar c L$.}
 \label{fig1}
\end{figure}

Fig. \ref{fig1} shows the interaction energy difference between
the eccentric and concentric configurations as a function of the
dimensionless eccentricity $\delta$. There are two sets of parameters used,
namely for $\alpha=b/a=2.0$ and $\alpha=2.5$. We have used 
$\epsilon_1=2$, $\epsilon_2=5$ and $\epsilon_3=50$ in all cases, varying
the relative order among them: $\epsilon_1 < \epsilon_2 < \epsilon_3$ and
$\epsilon_1 > \epsilon_2 > \epsilon_3$ for the positive curves, while 
$\epsilon_1 < \epsilon_2$, $\epsilon_3 < \epsilon_2$ and $\epsilon_1 > \epsilon_2$, 
$\epsilon_3 > \epsilon_2$ for the negative ones. 
The plot clearly shows the change of the sign of the energy
depending on the relative order among the dielectric constants,
and also the unstable/stable equilibrium position at the concentric
configuration.

\begin{figure}[ht]
\centering
\includegraphics[width=8.9cm]{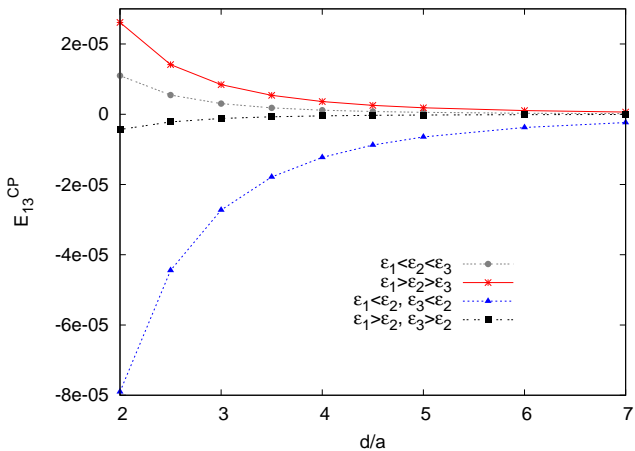}
\caption{(Color online). Exact Casimir interaction energy for the cylinder-plane
configuration as a function of the minimum distance between the
cylinder and the plane. In the upper curves, we show the
energy for the cases in which $\epsilon_1 < \epsilon_2 < \epsilon_3$
or $\epsilon_1 > \epsilon_2 > \epsilon_3$, and the force is
repulsive. The opposite case is shown in the lower part of the plot.
In these cases $\epsilon_1 > \epsilon_2$, $\epsilon_3
> \epsilon_2$ or $\epsilon_1 < \epsilon_2$, $\epsilon_3
< \epsilon_2$, and therefore the force is attractive. 
Energy is plotted in units of $\hbar c L$.}
\label{fig2}
\end{figure}

In order to analyze the dependence of the sign of the force in the cylinder-plane geometry on the
relation between $\epsilon_1$, $\epsilon_2$, and $\epsilon_3$ we
show in Fig.\ref{fig2} the exact Casimir interaction energy a function of the minimum distance
between the cylinder and the plane $d=H-a$.  Again, we plot different
orderings of the dielectric constants. It is easy to note that for
a decreasing or increasing ordering of the dielectric constants, we get
a repulsive force, while in any other case the force is attractive.

\begin{figure}[ht]
\centering
\includegraphics[width=8.9cm]{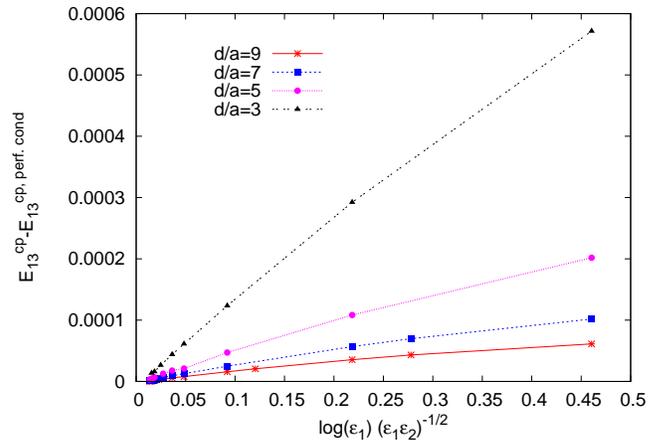}
\caption{(Color online). Exact Casimir interaction energy difference between the
dielectric cylinder-plane and the perfect conductors configurations.
As can be seen in Eq.(\ref{cylinder-plane-large-eps1}), the
convergence to the perfect conductors case is ruled by a $\ln
\epsilon_1 /\sqrt{\epsilon_1}$ coefficient, making it slower than
the naively expected from a simple  $1/\epsilon_1$ decay. In this
plot we have set $\epsilon_2 = 1$. Energy is plotted in units of $\hbar c L$.} \label{fig3}
\end{figure}

Finally, in Fig. \ref{fig3} we numerically check
Eq.(\ref{cylinder-plane-large-eps1}) with $\epsilon_3\to\infty$. In this plot we show the
difference between the interaction energy for the scalar dielectric
example and the perfectly conducting case, as a function of $\log
\epsilon_1 /\sqrt{\epsilon_1}$ (different values of the distance to
the plane are considered).

\section{Conclusions}

We have derived the exact expression for the scalar Casimir interaction energy between media-separated eccentric cylinders, and obtained the cylinder-plane
result as a particular case. In analogy with the well-known electromagnetic Casimir-Lifshitz interaction energy between fluid-separated planes, our results show
that the force in the two non-planar geometries studies can be repulsive or attractive depending on the relative strength of the permittivities of the three intervening
media over a broad range of frequencies.  We have presented both analytical and numerical calculations to prove that this is indeed
the case regardless the value of the radii of the cylinders and of the eccentricity.

We have considered the case of a quantum scalar field. The analogous calculation for the full electromagnetic field is much more involved. The reason is that, unlike the case of perfect reflectors, TE and TM modes do not decouple for finite values of the permittivities, and 
this fact introduces algebraic complications in the derivation of the exact 
formula presented in Section II. We expect to
analyze this issue in a future publication.
 

\section{Acknowledgments} 

We thank J. Etcheverry and F. Intravaia for useful discussions. The work of DARD was funded by DARPA/MTO's
Casimir Effect Enhancement program under DOE/NNSA Contract
DE-AC52-06NA25396. FCL and FDM were supported by UBA, CONICET,
and ANPCyT, Argentina. PIV acknowledges support of UNESCO LOREAL For Women 
in Science Programme.


\end{document}